\documentclass[10pt,twocolumn,letterpaper]{article}

\usepackage{iccv}
\usepackage{times}
\usepackage{epsfig}
\usepackage{graphicx}
\usepackage{amsmath}
\usepackage{amssymb}
\usepackage{tabularray}
\usepackage{booktabs}
\usepackage{tabularx}
\usepackage{float}
\usepackage{siunitx}
\usepackage[table]{xcolor}
\usepackage{hhline}
\usepackage[margin=2cm]{geometry}
\usepackage{adjustbox}
\usepackage{multirow}
\usepackage{array}
\usepackage{boldline} 
\usepackage{makecell}
\usepackage{xcolor}
\usepackage{pifont}
\newcommand{\xmark}{\ding{55}}

\usepackage[pagebackref=true,breaklinks=true,letterpaper=true,colorlinks,bookmarks=false]{hyperref}

\iccvfinalcopy 

\def\iccvPaperID{****} 

\begin{document}

\title{Implicit Neural Representation in Medical Imaging: A Comparative Survey}

\author{Amirali Molaei$^{1}$ \and Amirhossein Aminimehr$^{1}$ \and Armin Tavakoli$^{1}$ \and Amirhossein Kazerouni$^{2}$
\and Bobby Azad$^3$ 
\and Reza Azad$^4$ \and Dorit Merhof$^{~5,6,}$\thanks{Corresponding author: Dorit Merhof,$\:$ Tel.: +49 (941) 943-68509,$\:$ ~ ~ E-mail: dorit.merhof@ur.de.} \and 
\begin{minipage}{\textwidth}
\begin{flushleft}
\footnotesize$^1$\emph{School of Computer Engineering, Iran University of Science and Technology, Tehran, Iran} \\ 
\footnotesize$^2$\emph{School of Electrical Engineering, Iran University of Science and Technology, Tehran, Iran} \\
\footnotesize$^3$\emph{South Dakota State University, Brookings, South Dakota, USA} \\ 
\footnotesize$^4$\emph{Faculty of Electrical Engineering and Information Technology ,RWTH Aachen University, Aachen, Germany} \\ \footnotesize$^5$\emph{Institute of Image Analysis and Computer Vision, Faculty of Informatics and Data Science, University of Regensburg, Regensburg, Germany} \\
\footnotesize$^6$\emph{Fraunhofer Institute for Digital Medicine MEVIS, Bremen, Germany}
\end{flushleft}
\end{minipage}
}


\maketitle
\ificcvfinal\thispagestyle{empty}\fi



\begin{abstract}
Implicit neural representations (INRs) have gained prominence as a powerful paradigm in scene reconstruction and computer graphics, demonstrating remarkable results. By utilizing neural networks to parameterize data through implicit continuous functions, INRs offer several benefits. Recognizing the potential of INRs beyond these domains, this survey aims to provide a comprehensive overview of INR models in the field of medical imaging. In medical settings, numerous challenging and ill-posed problems exist, making INRs an attractive solution. The survey explores the application of INRs in various medical imaging tasks, such as image reconstruction, segmentation, registration, novel view synthesis, and compression. It discusses the advantages and limitations of INRs, highlighting their resolution-agnostic nature, memory efficiency, ability to avoid locality biases, and differentiability, enabling adaptation to different tasks. Furthermore, the survey addresses the challenges and considerations specific to medical imaging data, such as data availability, computational complexity, and dynamic clinical scene analysis. It also identifies future research directions and opportunities, including integration with multi-modal imaging, real-time and interactive systems, and domain adaptation for clinical decision support. To facilitate further exploration and implementation of INRs in medical image analysis, we have provided a compilation of cited studies along with their available open-source implementations on \href{https://github.com/mindflow-institue/Awesome-Implicit-Neural-Representations-in-Medical-imaging}{GitHub}. Finally, we aim to consistently incorporate the most recent and relevant papers regularly.

\end{abstract}

\section{Introduction}

\begin{figure*}
    \centering
    \includegraphics[width=1.6\columnwidth]{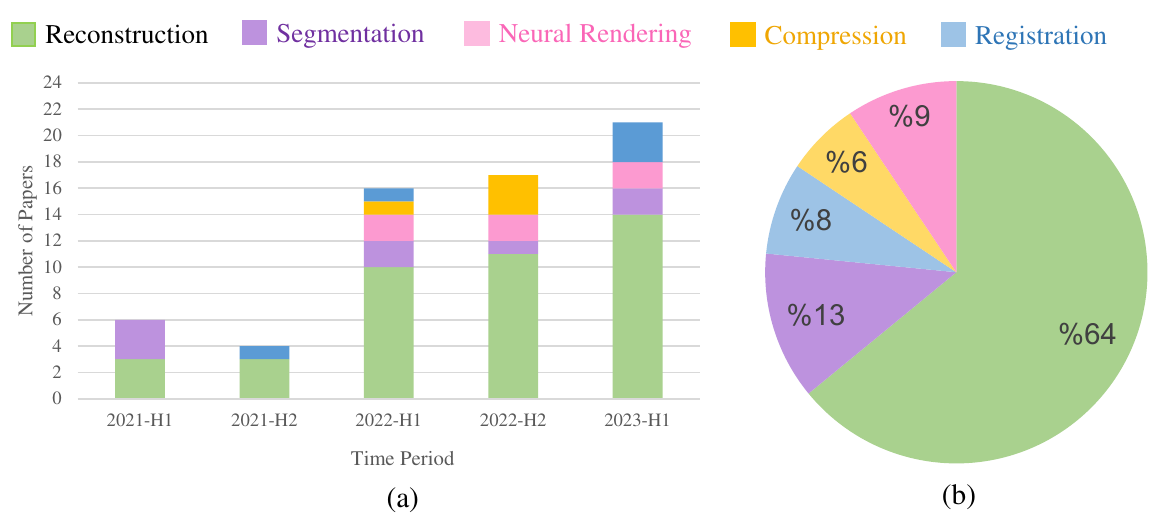}
    \caption{Chart (a) visually displays the relative proportions of published papers according to their application, and chart (b) illustrates the count of published papers based on INR design in medical tasks during different time periods (with "H" indicating the first or second half of the year). The assessment of the statistics is based on a sample of 64 research papers published during the years 2021 to 2023.}
    \label{fig:stats}
\end{figure*}

Knowledge representation is one of the fundamental pillars of artificial intelligence (AI). Its importance stems from its significant impact on the success of machines in learning many AI-related tasks, such as classification and decision-making. Humans construct task-specific representations to facilitate their interaction with the world around them \cite{niv2019learning}. Consequently, a substantial amount of AI research involves proposing algorithms and models that mimic this human cognition process to solve machine learning tasks with a degree of performance that equals, or even surpasses, human capabilities~\cite{silver2018general,azad2022medical,azad2023advances,aminimehr2023tbexplain}.

Our visual world can be represented continuously, which is a fundamental principle in fields such as computer vision. Data obtained through observation and sensing manifests in various forms, including images and audio. Conventional approaches to encoding input signals as representations typically follow an explicit paradigm, where the input space is discretized or partitioned into separate elements (\eg, point clouds, voxel grids, and meshes). However, in recent years, an alternative approach to representation, known as implicit representations, has gained popularity due to its efficient memory usage \cite{hao2022implicit}. Unlike explicit (or discrete) representations that directly encode the features or signal values, implicit representations are defined as a generator function that maps input coordinates to their corresponding value within the input space. 

In computer vision, the quality of representing an image signal holds central importance. Deep neural networks have emerged as the de facto tools for complex tasks across various AI domains, particularly in computer vision, owing to their remarkable representation learning ability \cite{kazerouni2023diffusion,aminimehr2023entri}. As a result, there has been an exploration of leveraging their capacity to function as implicit functions, yielding promising results \cite{sitzmann2020implicit,xie2022neural}. Within this context, a Multi-Layer Perceptron (MLP) is trained to parameterize the signal of interest, such as an image or shape, utilizing coordinates as input. The objective is to predict the corresponding data values at those coordinates. Thus, the MLP serves as an \textbf{Implicit Neural Representation} function that encodes the signal's representation within its weights. For instance, in the case of image signals, feeding the MLP with pixel coordinates leads to the generation of its RGB value as the output.

These implicit neural functions find extensive application in tasks such as image generation, super-resolution, 3D object shape reconstruction, and modeling complex signals \cite{mescheder2019occupancy,chabra2020deep,mildenhall2021nerf,wu2021irem}. The utilization of MLPs for image and shape parameterization offers several advantages. Firstly, they are resolution-agnostic as they operate in the continuous domain. This characteristic enables them to generate values for coordinates in-between pixel or voxel-wise grids, thereby facilitating vision tasks. Secondly, the memory requirements for representing the signal are not limited by its resolution. Consequently, MLP models exhibit enhanced memory efficiency while remaining effective, especially when compared to the grid or voxel representations. The memory demand scales according to the complexity of the signal itself. Thirdly, MLPs assist INRs in mitigating potential locality biases that could hinder performance, unlike convolutional neural networks (CNNs). Finally, MLP models are differentiable, endowing them with significant adaptability across a wide range of applications. Their weights can be adjusted using gradient-based techniques, allowing for versatility in handling different tasks \cite{chen2019learning,park2019deepsdf,sitzmann2020implicit,liu2020dist,xie2022neural,tancik2020fourier,saragadam2023wire,zhang2023dynamic}.

\begin{figure*}[t]
 \centering
 \includegraphics[width=0.9\textwidth]{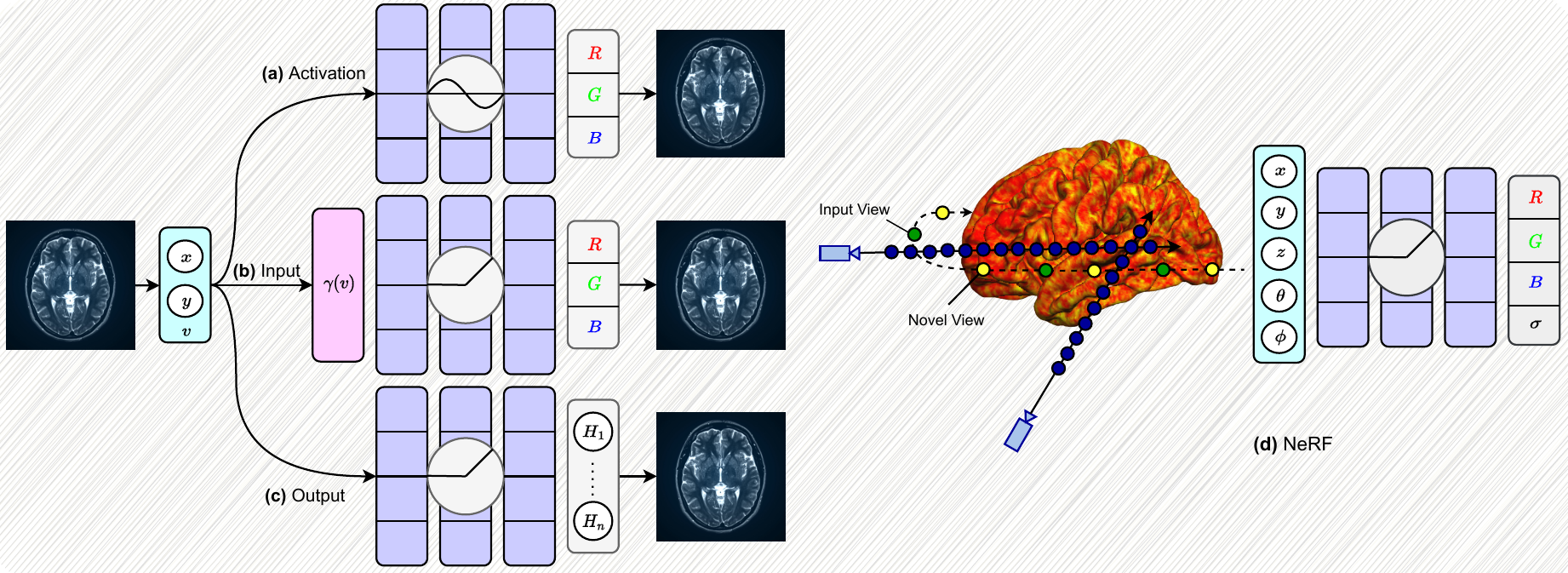}
 \caption{The figure illustrates various modifications to alleviate the spectral bias problem in INRs, provides an overview of their underlying principles, and introduces NeRF as an additional background method, as discussed in \autoref{sec:background}.}
 \label{fig:bg}
\end{figure*}

To address the challenges posed by the high cost of labeled data and limited memory resources, INRs have garnered increasing attention from the medical community, as evident from the exponential growth in research papers dedicated to this domain (\autoref{fig:stats}). This surge of interest in INR within the medical imaging community has resulted in a multitude of applications across diverse medical imaging scenarios. In recent years, their predominant usage has been in enhancing resolution and synthesizing missing information. They offer a solution to alleviate the burden of collecting labeled data in medical domains by eliminating the need for training data and explicit labels. Instead, they leverage available measurements or signals without the requirement of labeled data for each instance, enabling the reconstruction of 3D anatomical structures or generation of 2D scans \cite{sun2021coil,reed2021dynamic,shen2022nerp}. For instance, in MRI, super-resolution techniques can enhance the spatial resolution of images, providing clearer visualization of structural features and diagnostic information \cite{mao2022mri,wu2022arbitrary,wu2021irem}. Moreover, they can be employed in synthesis and inverse problems, such as reconstructing CT and MRI data from projection and frequency domains while reducing radiation exposure \cite{sun2021coil,zang2021intratomo,reed2021dynamic,shen2022nerp,xu2023nesvor}. Additionally, INRs find utility in neural rendering to model complex relationships in scenes, enabling detailed visualizations of anatomical structures or aiding in robotic surgery by reconstructing deformable surgical scenes \cite{corona2022mednerf,wang2022neural}.







To present a comprehensive review of these emerging architectures, this paper provides an overview of their core principles and diverse applications, as well as their advantages and limitations. To the best of our knowledge, this is the first survey paper that covers the application of INR in medical imaging and sheds light on new directions and research opportunities, serving as a roadmap and systematic guide for researchers. We also aspire to generate increased interest within the vision community to delve into the exploration of implicit neural representations in the medical domain.
Our key contributions are as follows:

\noindent$\bullet$ We conduct a systematic and comprehensive review of the applications of INR in the field of medical imaging. We analyze and compare state-of-the-art (SOTA) approaches for each specific task.

\begin{figure*}[t]
 \centering
 \includegraphics[width=0.92\textwidth]{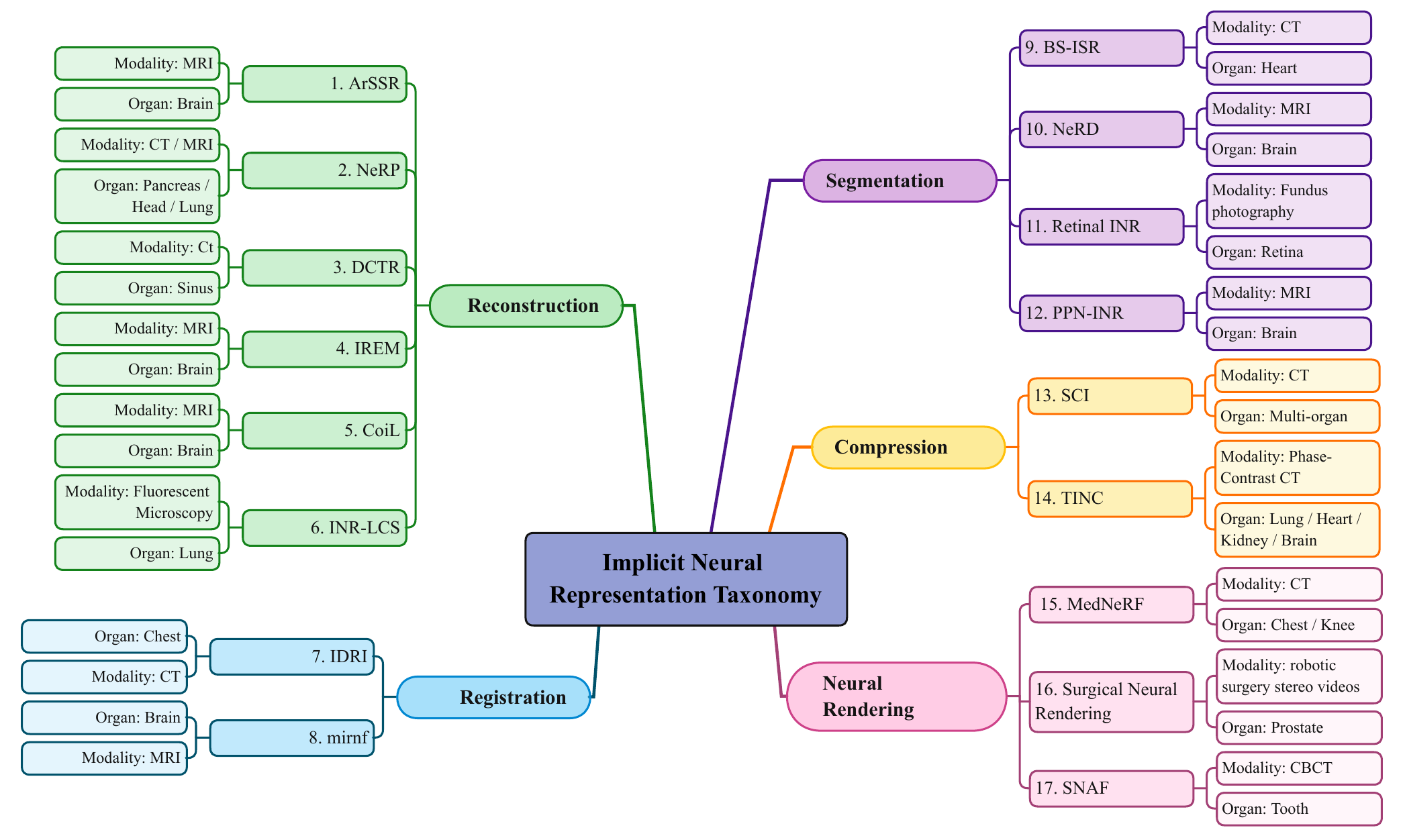}
 \caption{The taxonomy subsections delineate six distinct sub-fields in medical imaging: (1) Reconstruction, (2) Segmentation, (3) Registration, (4) Compression, and (5) Neural Rendering. We use the numbering of the methods in ascending order and provide the reference for their paper as follows: 1.~\cite{wu2022arbitrary}, 2.~\cite{shen2022nerp}, 3.~\cite{reed2021dynamic}, 4.~\cite{wu2021irem}, 5.~\cite{sun2021coil}, 
 6.~\cite{wiesner2022implicit},
 7.~\cite{wolterink2022implicit}, 
 8.~\cite{sun2022mirnf},  
 9.~\cite{barrowclough2021binary}, 
 10.~\cite{zhang2021nerd}, 
 11.~\cite{gu2022retinal}, 
 12.~\cite{li2023direct}, 
 13.~\cite{yang2022sci}, 
 14.~\cite{yang2023tinc}, 
 15.~\cite{corona2022mednerf}, 
 16.~\cite{wang2022neural},
 17.~\cite{fang2022snaf}.}
 \label{fig:taxonomy}
\end{figure*}

\noindent$\bullet$ We provide a detailed and organized taxonomy, as illustrated in \autoref{fig:taxonomy}, which allows for a structured understanding of the research progress and limitations in different areas.

\noindent$\bullet$ Additionally, we discuss the challenges and open issues associated with INR in medical imaging. We identify new trends, raise important questions, and propose future directions for further exploration.

\noindent\textbf{\emph{Search Strategy}}. To conduct a comprehensive search, we utilized DBLP, Google Scholar, and Arxiv Sanity Preserver, employing customized search queries tailored to retrieve relevant scholarly publications. Our search queries consisted of keywords such as \texttt{(INR $|$ implicit neural representation $|$ medical $|$ {Task})}, \texttt{(INR $|$ medical $|$ Neural rendering)}, and \texttt{(INR $|$ medical $|$ NeRF)}. Here, \textbf{{Task}} refers to one of the applications covered (\autoref{fig:taxonomy}).
To ensure the selection of relevant papers, we conducted a meticulous evaluation based on factors such as novelty, contribution, and significance. Priority was given to papers that were pioneering in the field of medical imaging. Subsequently, we selected papers with the highest rankings for further examination. 


\section{Background}\label{sec:background}
Implicitly representing signals with neural networks has gathered pace in recent years. Instead of parametrizing signals with discrete representations such as grids, voxels, point clouds, and meshes, a simple MLP can be learned to continuously represent the signal of interest as an implicit function $\Psi: \mathrm{x}\rightarrow \Psi{(\mathrm{x})}$, mapping their spatial coordinates $\mathrm{x} \in \mathbb{R}^M $ from $M$ dimensional space to their corresponding $N$ dimensional value $\psi{(\mathrm{x})} \in \mathbb{R}^N$ (\eg, occupancy, color, etc.). While INRs have shown promising, they can fail to encode high-frequency details compared to discrete representations, leading to a suppressed representation quality. Rahaman \etal \cite{rahaman2019spectral} have made significant strides in uncovering limitations within conventional ReLU-based MLPs and their ability to represent fine details in underlying signals accurately. These MLPs have shown a propensity to learn low-frequency details, leading to a phenomenon known as spectral bias in piece-wise linear networks. In order to address this issue, several approaches have been explored to redirect the network's focus toward capturing high-frequency details and effectively representing the signal with finer-grained details. To enhance the representation of the input signal, three avenues can be pursued within an MLP framework based on its structure. Firstly, one can consider changing the \textbf{\textit{input}} type by mapping it to a higher-dimensional space to enable the network to capture more intricate details within the signal. Secondly, another approach involves replacing the ReLU activation function with a new \textbf{\textit{activation function}} that better facilitates the learning of high-frequency components. Lastly, one can explore altering the \textbf{\textit{output}} of the MLP to a higher-dimensional space, where each node is responsible for reconstructing a specific part of the signal. 
In this section, we will provide a background based on the modifications that can be made to mitigate the spectral bias issue. Additionally, we will cover a neural volume rendering model called NeRF \cite{mildenhall2021nerf} as a pioneering approach to bridge implicit representations and novel view synthesis. \autoref{fig:bg} illustrates the overview of our proposed background. 

\subsection{Input}
The conventional approach in INR treats the spatial coordinate of each element in the signal, such as pixels in an image, as the input to an MLP. However, this approach tends to learn low-frequency functions, limiting its ability to effectively represent complex signals. To address this limitation, recent progress suggests using a sinusoidal mapping of the Cartesian coordinates to a higher dimensional space, which enables the learning of high-frequency details more effectively~\cite{tancik2020fourier}:

\begin{enumerate}
    \item \textbf{Basic:} $\gamma(\mathbf{v}) = [\text{cos}(2\pi\mathbf{v}, \text{sin}(2\pi\mathbf{v})]^T.$
    \item \textbf{PE:} 
    $\gamma(\mathbf{v}) = [..., \text{cos}(2\pi\sigma^{j/m}\mathbf{v}, \text{sin}(2\pi\sigma^{j/m}\mathbf{v}), ...]^T$ for $j = 0, ... , m-1$. \textbf{PE} denotes Positional Encoding, and the scale $\sigma$ is determined for individual tasks and datasets through a process of hyperparameter sweep.
    \item \textbf{Gaussian:} $\gamma(\mathbf{v}) = [\text{cos}(2\pi\mathbf{B}\mathbf{v}, \text{sin}(2\pi\mathbf{B}\mathbf{v}]^T,$ where the variable $\mathbf{v}$ represents the signal coordinates, while $\mathbf{B}$ is a random Gaussian matrix, where each entry is independently sampled from a normal distribution $\mathcal{N}(0, \sigma^{2})$. Similarly, the scale $\sigma$ is selected through a hyperparameter sweep for each task and dataset.
\end{enumerate}

These encoding processes are known as Fourier features mapping.
\subsection{Activation Function}
In general, the intuition behind activation functions is to apply non-linearity to the neural network. As for implicit representations, nonlinearities can be either periodic or non-periodic. However, non-periodic functions, such as $ReLU$ or $tanh$, are not conducive to the effective learning of high-frequency signals. To handle this issue, Sinusoidal Representation Networks (SIRENs) \cite{sitzmann2020implicit} utilize $sine$ as the activation function of the MLP to parametrize complex data:
\begin{equation}
\begin{aligned}
  &\mathbf{\Psi} (\mathbf{x})=\mathbf{W}_n (\psi _{n-1} \circ \psi _{n-2} \circ \dots \psi _0) (\mathbf{x})+\mathbf{b_n} , \\ &\mathbf{x_i} \mapsto \psi _{i}(\mathbf{x_i})=\mathrm{sin}(\mathbf{W_i} \mathbf{x_i} + \mathbf{b_i}),
  \label{eq:fourierfeatures}
\end{aligned}
\end{equation}
where $\psi _i$ indicates the $i^{th}$ layer of the neural network, $\mathbf{x}$ is the signal of interest, $\mathbf{W}$ and $\mathbf{b_i}$ represents the weight matrix and bias, respectively. 
The use of sine as the activation boils down to its derivative being a shifted sine (cosine), which enables the network to efficiently parametrize higher-order derivatives, such as the image Laplacian or Helmholtz equation. Furthermore, the sine function helps to effectively represent signals containing high-frequency details. 
SIREN authors suggest a unique initialization technique to prevent vanishing gradients in traditional activation functions. They initialize weights of each layer as $W\sim U(\frac{-c}{\sqrt{n}}, \frac{c}{\sqrt{n}})$, where $W$ is the weight of the layers, $U(.)$ is a uniform distribution, $c$ denotes a constant that controls the range of the weight values and $n$ is the number of inputs neurons.

\subsection{Output}
Target signals such as images and audio typically exhibit local structure and dependencies among neighboring elements, which can be effectively utilized to enhance ReLU networks during training. Aftab \etal \cite{aftab2022multi} introduce a multi-head network architecture where the main body learns global features of the signal, while the output layer consists of multiple heads. These heads reconstruct separate parts of the signal and learn its local features. For instance, in the case of images, they divide the image into equal grid cells. Each cell is then processed by an MLP in the main body to capture global features, and the output sparse nodes reconstruct the details of each cell individually. This approach aims to reduce the network's bias towards low-frequency components by exploiting the inherent properties of the target signal. Therefore, changing the output to a higher dimensional space can effectively alleviate the issue of spectral bias.
\subsection{NeRF}
Neural Radiance Fields (NeRFs) \cite{mildenhall2021nerf} unites INRs with volume rendering by using fully connected MLPs to implicitly represent scenes and objects with the goal of novel view synthesis. The objective of novel view synthesis is to develop a system that can generate novel viewpoints of an object from any direction by observing a few images of that particular object. The process is defined as:
\begin{equation}
\begin{aligned}
F_\theta (\mathrm{x}, \mathrm{d}) \longrightarrow (\mathrm{c}, \sigma),
  \label{eq:nerfmappingfunction}
\end{aligned}
\end{equation}
where $\mathrm{x}$ indicates the 3D location $(x, y, z)$, $\mathrm{d}$ represents the 2D vector of viewing direction $(\theta,\phi)$, $\mathrm{c}$ denotes the color values $(r, g, b)$, and $\mathrm{\sigma}$ is the volume intensity. The primary idea is to overfit an implicit function on the training data such that, given $(x, y, z)$ as the spatial coordinates and ($\theta, \phi$) as the viewing direction, the network outputs the color and volume density of the particular location. Unlike SIREN~\cite{sitzmann2020implicit}, the NeRF architecture is equipped with ReLU as its activation function, but similar to Fourier features \cite{tancik2020fourier}, adopts a positional encoding approach to map coordinates to higher dimensions as follows: \begin{equation}
\begin{aligned}
\gamma(p) = (sin(2^0\pi p) , cos(2^0\pi p) , \dots ,\\ sin(2^{L-1}\pi p) , cos(2^{L-1}\pi p)),
  \label{eq:positionalencoding}
\end{aligned}
\end{equation}
where $p$ can be each of the coordinate or viewing direction components. NeRF's architecture is designed in a two-stage matter to obtain density and color values as follows:
\begin{flalign}
&\mathrm{First\ Stage}: \mathrm{\sigma}, \mathrm{h} = MLP(\mathrm{x})&\\
&\mathrm{Second\ Stage}: \mathrm{c} = MLP(concat[\mathrm{h}, \mathrm{d}])&
\label{eq:stages}
\end{flalign}
where in the first stage the 3D coordinate $\mathrm{x}$ is passed through the MLP to obtain the density $\mathrm{\sigma}$ and a feature representation $\mathrm{h} \in \mathbb{R}^N$ ($N$ equals 256 in original implementation), and in the second stage, $\mathrm{h}$ is input to the MLP to output the color values $\mathrm{c}$ (the original implementation \cite{mildenhall2021nerf} adopts identical MLPs to accomplish this). Finally, volume rendering \cite{kajiya1984ray} is utilized to generate novel views by tracing camera rays through every pixel of the target synthesized image.

\begin{figure}
  \centering
  \includegraphics[width=\columnwidth]{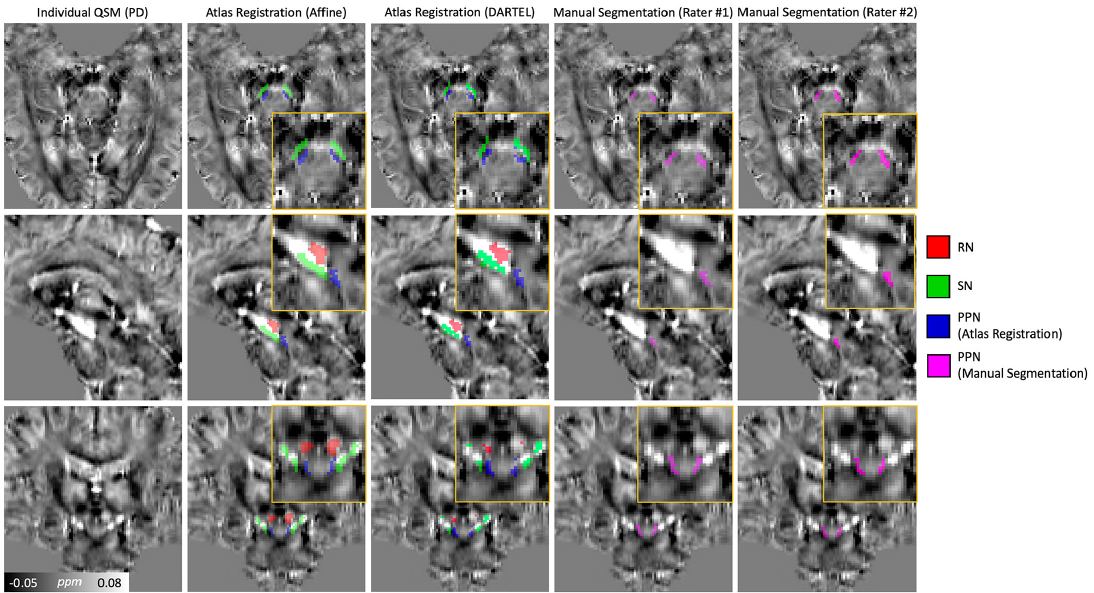} %
  \caption{Comparison of the aligned Pedunculopontine Nucleus (PPN) region (blue) and manually segmented PPN region (lavender) by radiologists in a Parkinson's disease patient~\cite{li2023direct}. The brain planes (axial, sagittal, and coronal) show the results of affine and Diffeomorphic (DARTEL) registration, aligning the myelin staining histological atlas and the enhanced atlas generated through INR-based super-resolution. Expert radiologists manually segmented the last two columns to validate the INR-based method.} 
  \label{fig:case_study}
\end{figure}


\section{Clinical Importance}\label{sec:importance}

Thanks to the memory and data efficiency of INRs, they are widely utilized in numerous medical imaging tasks. One of the most significant challenges in automated medical imaging is the collection of ground truth annotated data from reliable sources, such as clinicians and medical professionals~\cite{hesamian2019deep,altaf2019going,kazerouni2023diffusion}. This process is painstaking, expensive, time-consuming, and requires significant effort. Unlike simple scenes that can be easily recognized and labeled (\eg, categorizing an indoor scene), annotating medical images should be performed by medical professionals and clinicians. This reliance on experts, coupled with privacy concerns and the need for patient authorization, creates a major bottleneck in the annotation process for medical imaging. INRs offer significant advantages in various applications without the need for external training annotations~\cite{sun2021coil,park2021hypernerf,reed2021dynamic,shen2022nerp,wolterink2022implicit}. For example, in the field of medical imaging, INRs are particularly beneficial for tasks like super-resolution. During the medical imaging procedures like CT scans, PET scans, MRI, and ultrasound, patient movement can also cause motion artifacts and result in blurred images or poorly defined structures, especially in upper abdominal regions such as the chest which are negatively affected by patient motion. Additionally, in Cone Beam Computed Tomography (CBCT), which is commonly used in dental and maxillofacial imaging, slow imaging speed combined with patient movement can result in motion artifacts and lead to poorly defined structure boundaries~\cite{sonke2005respiratory,fang2022snaf,zhang2023dynamic}. Furthermore, obtaining high-quality MRI scans poses a challenge due to longer scan times~\cite{wu2022arbitrary,wu2021irem}. Conventional approaches are not suitable for effectively handling these issues through super-resolution or image reconstruction as they are not resolution-agnostic and require notable amounts of data. However, INRs can address super-resolution more effectively by considering inputs from the continuous coordinate domain and being resolution-agnostic. 

Implicit neural models are also widely used in biomedical applications, particularly in solving inverse imaging problems~\cite{zang2021intratomo,reed2021dynamic,shen2022nerp,shen2022nerp}. These problems involve learning the structure of an object (organ of interest in medical cases) from observations or measurements. Using INRs, it becomes possible to reconstruct CT or MRI scans directly from the sensor domain. Moreover, they can even facilitate the tracking of tissue progression by incorporating prior scans from earlier time steps, subsequently reconstructing the updated scan for the current time.

In practical applications, the reconstruction of images from sparsely sampled data plays a crucial role. This need arises in various domains, including medical imaging, where it has proven particularly valuable in specific applications such as reducing radiation dose in CT imaging and accelerating MRI scans~\cite{sun2021coil,reed2021dynamic,shen2022nerp}. Notably, Single Image Super-Resolution (SISR) techniques have attracted considerable attention due to their potential to restore a high-resolution (HR) image solely based on a low-resolution (LR) input~\cite{wu2021irem,mao2022mri}. The ability of SISR methods to enhance image details and fidelity has significant implications for improving diagnostic accuracy and aiding medical professionals in their decision-making processes. This cannot be accomplished using convolution-based techniques as they are trained exclusively for particular up-scaling tasks. Additionally, the time-consuming process of retraining them for a new up-scaling task hinders their usefulness in clinical applications \cite{wu2022arbitrary}.

INRs have also found application in assisting robotic surgery~\cite{schmidt2022recurrent,schmidt2022fast,wang2022neural,zhang2023implicit,alblas2023implicit}. The integration of INRs within robotic surgical systems allows for enhanced perception and understanding of the surgical environment. By leveraging INRs, robotic surgical systems can better interpret intraoperative images, providing real-time feedback and guidance to the surgeon. This can assist in accurate tissue segmentation, localization of critical anatomical structures, and precise surgical tool manipulation.

To underscore the practical applicability of INRs in addressing real-world challenges, it is crucial to highlight the validation of these models through human expertise. By corroborating the findings of INRs with the insights of medical professionals, we can establish the reliability and effectiveness of these novel diagnostic tools. This notion is exemplified by a compelling study that utilized the expertise of radiologists to validate the outcomes. In the depicted case study (\autoref{fig:case_study}), the aligned Pedunculopontine Nucleus (PPN) and the manually segmented PPN region by two radiologists are compared for a patient with Parkinson's disease~\cite{li2023direct}. The study focuses on enhancing the visibility and localization of the PPN, a deep brain structure crucial for Parkinson's disease treatment, through a combination of a Quantitative Susceptibility Mapping Atlas (QSM) and an INR network. The INR-based process significantly improves the spatial resolution of the atlas, effectively overcoming limitations and minimizing artifacts and blurring effects. As a result, it allows for better delineation of the specific region (PPN) on the atlas, indicating the usability and effectiveness of this approach in clinical settings.

In conclusion, INRs models offer significant advantages in medical imaging tasks, addressing challenges such as the lack of annotated data and artifacts in scans. We believe that research can exploit experts to validate their method's practicality, therefore better demonstrating the usefulness of INRs.
In conclusion, INRs have emerged as a valuable and adaptable tool in clinical settings, successfully tackling a diverse range of imaging challenges. Their widespread use is anticipated to continue to grow in the future, offering new possibilities for medical imaging research.


\section{Taxonomy}\label{sec:taxonomy}

In this section, we provide a taxonomy with a focus on the application of INRs in several medical imaging tasks to acquaint researchers with the notable operation and functionality of these models.

\subsection{Reconstruction}
Image reconstruction is a critical task in medical analysis, enabling professionals to obtain high-quality images for clinical applications. Many studies have explored the use of convolutional neural networks (CNNs) to learn a mapping function that transforms raw data into reconstructed images. However, this approach faces challenges, including the need for large-scale training datasets, instability in the presence of structural modifications, and difficulties in generalizing to diverse image modalities or anatomical locations \cite{antun2020instabilities}. Overcoming these obstacles is crucial to improve the reliability and applicability of image reconstruction in medical settings.
To use INR here, the task is conventially defined as an inverse problem in medical image reconstruction takes noisy or undersampled measurements of a medical image as input and aims to generate a reconstructed, complete image as output. The input can come from various imaging modalities such as CT, MRI, or ultrasound, and the incompleteness may be due to time constraints, reduced radiation exposure, or patient movement. The INR model learns to map the input measurements to the corresponding complete images, recovering the missing information and producing high-quality images that resemble the ground truth obtained from fully sampled acquisitions.

\begin{figure*}[ht]
  \centering
  \includegraphics[width=\textwidth]{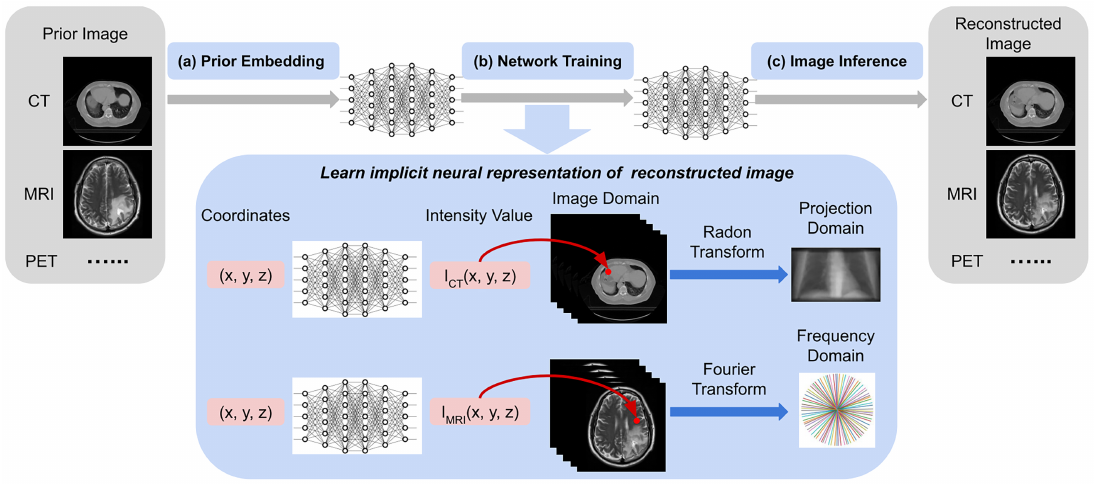}
  \caption{Overview of the NeRP framework (taken from \cite{shen2022nerp}). NeRP solves the reconstruction problem in three stages: (a) Embedding a prior medical image in an MLP by implicitly encoding it in its weights. (b) Adapting the MLP on the target image by minimizing the loss between the measurements of the prior and the target image. (c) Inferencing the actual target CT or MRI image from each of the coordinates.}
  \label{fig:NeRP}
\end{figure*}

To address the aforementioned challenges, numerous INR-based reconstruction methods have been developed in recent years. For instance, \textbf{NeRP} \cite{shen2022nerp} framework proposes to integrate implicit neural networks for reconstructing sparsely sampled medical images through three stages without demanding any training data. As illustrated in \autoref{fig:NeRP}, in the first stage, a neural network's weights are encoded with a CT image as prior knowledge. Next, the implicit network is optimized on sparsely sampled sinogram measurements to learn reconstruction. Finally, the network is applied to all associated spatial coordinates to generate the final reconstructed CT image. The effectiveness of NeRP in reconstructing tumor structural progression and its versatile applicability across various imaging modalities have been demonstrated through experiments conducted on 2D and 3D data, including CT clinical scans and brain tumor progression MRI data images. 

Additionally, Reed \etal~\cite{reed2021dynamic} proposed a method (\textbf{DCTR}) for reconstructing dynamic, time-varying scenes using computed tomography (4D-CT). The INR is utilized to estimate a template reconstruction of the 3D volume's linear attenuation coefficients (LACs) in the scene, acting as a prior model that captures the spatial distribution of LACs. Here, the template refers to a representation or approximation of the scene's properties, specifically the LACs in the 3D volume. By using the INR, DCTR generates a template reconstruction of the LACs based on available CT measurements or sinograms through learning a mapping between coordinates $(x, y, z)$ and the template reconstruction of the LACs, which serves as a starting point for the overall reconstruction process. DCTR then employs a parametric motion field, a set of parameters describing how the template should be warped over time to account for scene motion. Finally, the warped template reconstruction is used to synthesize sinograms through a differentiable Radon transform, which is then compared to the actual sinogram to evaluate the accuracy of the reconstruction. The proposed method demonstrates robust reconstruction of images with deformable and periodic motion and is validated on their synthetic D4DCT~\cite{reed2021dynamic} dataset and the thoracic CT data~\cite{castillo2009framework}.


\subsection{Segmentation}
Medical image segmentation is a critical task in healthcare systems, aiding in disease diagnosis and treatment planning. Deep learning methods have shown promising results in achieving accurate segmentation results. However, these methods often suffer from computational inefficiency and difficulty in handling complex topologies~\cite{tang2019skeleton}. Complex topologies refer to intricate structural relationships and variations within medical images, such as lesions, tumors, and intricate vessel structures. 

To address these limitations, Barrowclough~\etal \cite{barrowclough2021binary} introduced a novel approach called \textbf{BS-ISR} that combines convolutional neural networks (CNNs) with INRs. INRs were specifically selected for their ability to handle complex, high-dimensional medical imaging data and capture intricate topologies. Instead of directly generating images, the model utilizes spline representations to capture geometric boundaries and structures. The authors also introduced new loss functions tailored for modeling implicit splines, utilizing binary inside-outside masks. Evaluation on the Congenital Heart Disease dataset~\cite{xu2020imagechd} demonstrated the superior performance of the model compared to other SOTA methods, as measured by the average volumetric test Dice score metric.
In another research, Gu \etal~\cite{gu2022retinal} proposed a self-distillation-based INR method for segmentation of retinal vessels for ocular disease diagnosis (\textbf{Retinal INR}). They utilized the Vision Transformer (ViT)~\cite{dosovitskiy2020image} to capture global dependencies in retinal images by treating the image as a sequence of patches rather than focusing solely on local features. The self-distillation method extracted key features for blood vessel segmentation. The primary benefit of the suggested approach lies in its ability to enhance the resolution of retinal images and magnify the finer details of capillaries through the use of INR. To ensure accurate results, they utilized an improved centerline dice (clDice) loss function to constrain blood vessel topology. The proposed model was evaluated on the Drive~\cite{staal2004ridge} and Chase~\cite{fraz2012ensemble} datasets, showcasing its superiority over non-INR methods in terms of segmentation accuracy, detection of detailed structures, and robustness to variations in image quality and content.

\subsection{Registration}

\begin{figure*}[t]
  \centering
  \includegraphics[width=\textwidth]{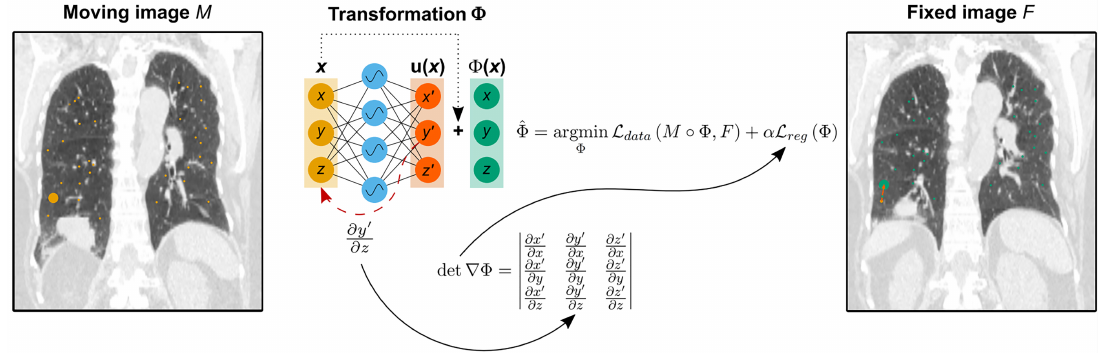}
  \caption{An illustration of the IDIR registration framework (taken from \cite{wolterink2022implicit}), which uses Implicit Neural Representations (INRs) in a multi-layer perceptron (MLP) to directly optimize the deformation function $\Phi(x) = u(x) + x$, which maps image coordinates to the deformation vector $u(x)$.}
  \label{fig:IDIR}
\end{figure*}

Medical image registration is a process that aligns multiple images, volumes, or surfaces within a common coordinate system to identify common areas. It requires learning a transformation function that geometrically aligns coordinates between a source and target image. Traditional methods often require complex, multi-step processes and assumptions about the nature of the transformations. However, INRs are capable to model the transformation function without external assumptions. This allows them to handle the variations and warping in the image in a smooth, coherent manner and at any image resolution, making them ideally suited for tasks such as image registration.

In this regard, Wolterink \etal \cite{wolterink2022implicit} proposed \textbf{IDIR}, that employs INR to model the transformation function with a SIREN-based design for deformable registration, which attempts spatial alignment of images to account for changes in the shape, position, or size of anatomical structures, such as organs, tumors, or other features of interest. As depicted in \autoref{fig:IDIR}, the transformation function $\phi(x) = u(x) + x$, which maps each coordinate $x$ in a fixed image to a coordinate in a moving image, is represented using the MLP. The MLP takes a continuous coordinate $x$ from the image domain as input and predicts a deformation vector $u(x)$. The addition of $u(x)$ and $x$ gives the output $\phi(x) = u(x) + x$. Moreover, the periodic activation function in the MLP allows for higher-order derivatives, enabling advanced regularization techniques for accurate and flexible image registration without relying on CNNs.
This model was tested on 4D chest CT registration using the DIR-LAB dataset \cite{castillo2009framework} and surpassed all deep learning-based methods without folding or training data needed. 

In another study, Sun \etal \cite{sun2022mirnf} developed \textbf{mirnf}, which can model both displacement vector fields and velocity vector fields, providing two different approaches for performing image registration. While displacement vector fields are used for deformable registration, velocity vector fields are employed for diffeomorphic registration, both utilizing INRs to model the transformations between the target and moving images. The network in registration based on velocity predicts velocity vectors $[v_{p_x},v_{p_y},v_{p_z}]$ from 3D coordinates in the target image. These vectors are integrated over time using a Neural ODE Solver to generate a deformation field by mapping each point in the target image to its corresponding deformed position in the moving image. By applying the deformation field to the target image, it can be aligned with the moving image. Alternatively, another approach trains an MLP to directly predict displacement vectors $[\phi_{p_x},\phi_{p_y},\phi_{p_z}]$ for each coordinate in the source image. These vectors describe how each point in the source image should be shifted or deformed to align with the target image. The target volume is deformed to match the source volume by applying these displacement vectors, which involves adding the displacement vector to each point's position in the target volume. The authors conducted experiments on two 3D MR brain scan datasets, Mindboggle101 \cite{klein2012101} and OASIS \cite{marcus2007open}, and found that INR achieves SOTA performance in terms of registration accuracy, optimization speed, and regularity compared to traditional methods.

\subsection{Compression}

\begin{figure*}[ht]
  \centering
  \includegraphics[width=\textwidth]{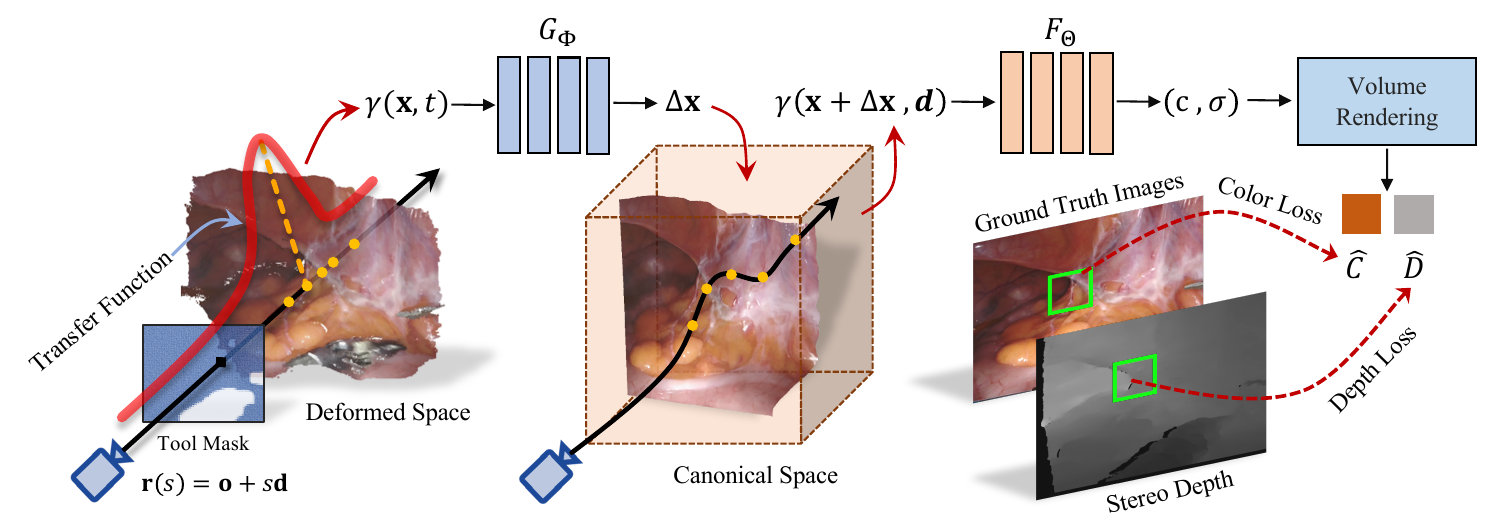}
  \caption{The Surgical Neural Rendering framework proposed by Wang \etal (figure taken from \cite{wang2022neural}). The surgical scenes are represented using a canonical radiance field $F_{\theta}(x, d)$ and a time-dependent displacement field $G_{\Phi}(x, t)$. Both models are designed using MLPs, where the canonical radiance field takes as input the spatial coordinates $\mathrm{x} \in \mathbb{R}^3$ and the unit view-in directions $d \in \mathbb{R}^3$, and the displacement field takes as input the space-time coordinates $(x, t)$. The output of these MLPs is the RGB colors $c(x, d) \in \mathbb{R}^3$ and space occupancy $\sigma(x) \in \mathbb{R}$ for the canonical radiance field and the displacement vector at point $x$ and time $t$ for the displacement field.}
  \label{fig:RoboticSurgery}
\end{figure*}

With increasing volumes of biomedical data, efficient compression methods are needed for storage, transmission, and secure sharing. While compression techniques for natural image/video data exist, they are not effective for biomedical data due to their unique characteristics. Biomedical data contains diverse tissue types, complex structures, and high-resolution details, which pose challenges for conventional compression techniques. In recent years, target-data-specific approaches like INR have shown promise in effectively compressing diverse visual data.


For instance, Yang \etal \cite{yang2022sci} presented a mathematical interpretation and adaptive partitioning for the design of an INR-based compressor called \textbf{SCI}. SCI partitions the data into blocks, and each block is compressed separately using an MLP network. The first layer is allocated a wide set of neurons to capture a broader range of frequencies, with a proportional reduction of layer size with increasing depth. This choice is based on the observation that increasing the depth of the network rather than the width (number of neurons) is more efficient to represent a larger range of frequencies or higher-order harmonics. To maintain high reconstruction fidelity, the allocation of parameters to blocks is done in accordance with the range of frequencies they cover. After the compression is completed for each block, the network parameters, including the weights and biases of the neural network that contains the learned representations and encoding information for that specific block, are serialized.
Using the HiP-CT dataset \cite{walsh2021imaging} as a testbed, Yang \etal found that their method outperformed conventional techniques (JPEG, H.264, HEVC), data-driven techniques (DVC, SGA+BB, SSF), and existing INR-based techniques (SIREN \cite{sitzmann2020implicit}, NeRF \cite{mildenhall2021nerf}, and NeRV \cite{sarkar2020neural}) on a wide variety of biological and medical data.

In another attempt to improve the compression fidelity of INR, a Tree-structured Implicit Neural Compression (\textbf{TINC}) was proposed \cite{yang2023tinc}. TINC uses MLPs to fit segmented local regions, and these MLPs are arranged in a tree structure to enable parameter sharing based on spatial distance. The parameter-sharing mechanism ensures a smooth transition between neighboring regions and eliminates redundancy, whether it exists locally or non-locally. The experiments on the HiP-CT dataset \cite{walsh2021imaging} demonstrate the superiority of TINC over conventional techniques. However, its limitations, similar to those of other INR-based methods, are its slower compression speed, although its decompression speed is high.

\subsection{Neural Rendering}

Neural rendering refers to a class of approaches that involve training a neural network to model the complex relationships between scene geometry, lighting, and details, which allows for the generation of novel views based on existing scenes. Implicit representations can be applied in the context of neural rendering of medical images, allowing for the creation of more detailed and accurate visualizations of complex anatomical structures and other medical data.

In 3D CT imaging, the long exposure time of patients to harmful ionizing radiation imposes a noticeable challenge. Consequently, to alleviate this problem, \textbf{MedNeRF} \cite{corona2022mednerf} proposes to incorporate GRAF \cite{schwarz2020graf} (which integrates NeRF~\cite{mildenhall2021nerf} with a CNN) to render CT projections from a single or multi-view X-rays. The intention behind GRAF boils down to NeRF struggling to handle complex scenes with large amounts of geometric complexity. To handle this limitation, the NeRF is trained to minimize the difference between the rendered and ground truth images, while the GAN~\cite{goodfellow2014generative} is trained to distinguish between the generated image and a ground truth image, and utilized to refine the NeRF outputs and improve image quality. The conducted evaluations of MedNeRF on X-ray chest and knee datasets demonstrate reconstruction improvements in terms of volumetric depth estimation compared to the neural radiance field methods.

The application of neural rendering for the purpose of reconstructing surgical scenes in 3D was first introduced by Wang \etal~\cite{wang2022neural}. As shown in the \autoref{fig:RoboticSurgery}, the proposed method (\textbf{Surgical Neural Rendering}) employs Implicit Neural Representations (INRs) to capture the dynamic and deformable nature of surgical scenes through a canonical radiance field and a time-dependent displacement field, represented using an MLP that maps coordinates and view-in directions to RGB colors and space occupancy. By making the volume rendering process differentiable, it becomes possible to backpropagate gradients through the rendering operations, allowing for end-to-end learning of the implicit neural fields, and enabling the optimization of these parameters to reconstruct the surgical scenes. To generate renderings for supervision, the approach utilizes differentiable volume rendering, where camera rays are shot into the scene, and the color and optical depth of each ray are evaluated using the volume rendering integral. Sampled points along the rays provide the necessary inputs to obtain color and space occupancy from the neural fields. The network parameters of the implicit neural fields are optimized to reconstruct the shapes, colors, and deformations of the surgical scene. This optimization is achieved by jointly supervising the rendered color and optical depth with ground-truth data.

\definecolor{silver}{RGB}{192,192,192}

\definecolor{lightgreen}{RGB}{184,244,184}
\definecolor{llightgreen}{RGB}{225,250,225}

		
\definecolor{lightpurple}{RGB}{213,172,213}
\definecolor{llightpurple}{RGB}{237,221,237}
		
\definecolor{lightpink}{RGB}{255,205,213}
\definecolor{llightpink}{RGB}{255,234,237}							

\definecolor{lyellow}{RGB}{255,232,141}
\definecolor{llyellow}{RGB}{255,244,197}	

\definecolor{skyblue}{RGB}{135,206,235}
\definecolor{lskyblue}{RGB}{205,235,247}

\newcommand{\MRT}[2]{\multirow{#1}{*}{#2}}

\newcolumntype{C}[1]{>{\arraybackslash}p{#1}}
\newcolumntype{M}[1]{>{\centering\arraybackslash}m{#1}}

\begingroup
\def\arraystretch{1.8}%
\begin{table*}[h]
    \centering
    \fontsize{6}{7.5}\selectfont
    \caption{A comparison of different proposed methods in different medical imaging task. The table is divided into different sections based on the imaging task, including Reconstruction, Segmentation, Neural Rendering, Compression, and Registration.}
    \label{tab:highlights}
    
    \resizebox{\textwidth}{!}{
        \begin{tabular}{V{4}c|c|c|c|c|c|ccV{4}}

            \hlineB{4}
            \rowcolor{silver}
            \textbf{Application} & \textbf{Method}  & \textbf{Input} & \textbf{Output}  & \makecell{\textbf{Activation} \\ \textbf{Function}} & \textbf{Modality} & \textbf{Highlight} & \textbf{Ref} \\
            \hlineB{4}

            \rowcolor{llightgreen}  
            \cellcolor{lightgreen} & 
            \multicolumn{1}{c}{NeRP} & 
            \multicolumn{1}{c}{$(x, y, z)$} & 
            \multicolumn{1}{c}{Intensity} & 
            \multicolumn{1}{c}{ReLU} & 
            \multicolumn{1}{c}{\Gape[0pt][2pt]{\makecell{MRI \\ CT }}} & 
            \multicolumn{1}{C{1.1\columnwidth}}{Reconstructs high-quality CT and MRI images from sparsely sampled measurements by embedding a prior image from an earlier scan, training the network to learn the neural representation of the target image, and inferring the trained network to generate the final reconstructed image.} & 
            \multicolumn{1}{lV{4}}{\cellcolor{white}\cite{shen2022nerp}}  \\
            \cline{2-8}
           
            \rowcolor{llightgreen} 
             \cellcolor{lightgreen} & 
             \multicolumn{1}{c}{DCTR} &  
             \multicolumn{1}{c}{$(x, y, z)$} & 
             \multicolumn{1}{c}{\Gape[0pt][2pt]{\makecell{Linear \\ Attenuation \\ Coefficient} }} & 
             \multicolumn{1}{c}{ReLU} & 
             \multicolumn{1}{c}{CT} & 
             \multicolumn{1}{C{1.1\columnwidth}}{
                 Calculating the loss between the new CT measurement and the measurement taken in the previous time step, then propagates the loss back to the MLP in order to remove noise from the new scan and reconstruct its CT image.} & 
             \multicolumn{1}{lV{4}}{\cellcolor{white}\cite{reed2021dynamic}}  \\ 
             \cline{2-8}
            
            \rowcolor{llightgreen} 
             \cellcolor{lightgreen} & 
             \multicolumn{1}{c}{IREM} & 
             \multicolumn{1}{c}{$(x, y, z)$} & 
             \multicolumn{1}{c}{Intensity} & 
             \multicolumn{1}{c}{ReLU} & 
             \multicolumn{1}{c}{MRI} & 
             \multicolumn{1}{C{1.1\columnwidth}}{Generates  3D MRI volumes by fitting implicit neural representations on sagittal, coronal, and axial planes of the brain to reconstruct the vacant voxels between them.} & 
             \multicolumn{1}{lV{4}}{\cellcolor{white}\cite{wu2021irem}}  \\ 
             \cline{2-8}
            
            \rowcolor{llightgreen} 
             \cellcolor{lightgreen} & 
             \multicolumn{1}{c}{ArSSR} & 
             \multicolumn{1}{c}{\Gape[0pt][2pt]{\makecell{$(x, y, z) $ \\ Image Embedding} }} & 
             \multicolumn{1}{c}{Intensity} & 
             \multicolumn{1}{c}{ReLU} & 
             \multicolumn{1}{c}{MRI} & 
             \multicolumn{1}{C{1.1\columnwidth}}{represents the 3D brain MRI surface using high-resolution (HR) and low-resolution (LR) image pairs. The LR image is processed through task-specific super-resolution CNN to extract features, which are concatenated with the HR image coordinates. This concatenated input is then used by a decoder MLP to generate the 3D voxel intensity.} & 
             \multicolumn{1}{lV{4}}{\cellcolor{white}\cite{wu2022arbitrary}}  \\ 
             \cline{2-8}

             \rowcolor{llightgreen}  
            \cellcolor{lightgreen} & 
            \multicolumn{1}{c}{\MRT{3}{INR-LCS}} & 
            \multicolumn{1}{c}{\Gape[0pt][2pt]{\MRT{3}{\makecell{$(x, y, z)$ \\ $\tau$ (Temporal variable) \\ $z$ (latent code)}}} }& 
            \multicolumn{1}{c}{\MRT{3}{SDF}} & 
            \multicolumn{1}{c}{\MRT{3}{Sine}} & 
            \multicolumn{1}{c}{\Gape[0pt][2pt]{\MRT{3}{\makecell{Fluorescent\\Microscopy}}}} & 
            \multicolumn{1}{C{1.1\columnwidth}}{$\bullet$ Uses an MLP to represent the signed distance function (SDF) of evolving cell shapes at any point given its spatial coordinates (x, y, z), a temporal parameter (t), and a conditioning latent code that provides additional information and influences the generated cell shapes. \newline
            $\bullet$ Can aid researchers in studying subjects such as cell division, migration, or modifications in cellular structures.} & 
            \multicolumn{1}{lV{4}}{\cellcolor{white}\cite{wiesner2022implicit}}  \\

            \cline{2-8}
            
            \rowcolor{llightgreen} 
             \multirow{-10}{*}{\cellcolor{lightgreen}\textbf{\scriptsize Reconstruction}}& 
             \multicolumn{1}{c}{CoiL} & 
             \multicolumn{1}{c}{($\theta, l$)} & 
             \multicolumn{1}{c}{Response ($r$)} & 
             \multicolumn{1}{c}{ReLU} & 
             \multicolumn{1}{c}{CT} & 
             \multicolumn{1}{C{1.1\columnwidth}}{By utilizing the viewing angle $\theta$ and the spatial location $l$ of the relevant detector on the sensor plane as inputs, an MLP learns a mapping to sensor responses (CT measurement) in order to represent the measurement field of the organ of interest.} & 
             \multicolumn{1}{lV{4}}{\cellcolor{white}\cite{sun2021coil}}  \\ 
             
             \hline


             
            \rowcolor{llightpurple}  
            \cellcolor{lightpurple} & 
            \multicolumn{1}{c}{BS-ISR} & 
            \multicolumn{1}{c}{$(x, y)$} & 
            \multicolumn{1}{c}{Spline Coefficient} & 
            \multicolumn{1}{c}{ReLU} & 
            \multicolumn{1}{c}{CT} & 
            \multicolumn{1}{C{1.1\columnwidth}}{Used a combination of INR and CNNs to model the segmentation boundary by mapping CT slice coordinates to spline coefficients.} & 
            \multicolumn{1}{lV{4}}{\cellcolor{white}\cite{barrowclough2021binary}}  \\
            \cline{2-8}

            \rowcolor{llightpurple}  
            \cellcolor{lightpurple} & 
            \multicolumn{1}{c}{NeRD} & 
            \multicolumn{1}{c}{$(d_t, d_r, d_b, d_l)$} & 
            \multicolumn{1}{c}{$(\mu, \Sigma)$} & 
            \multicolumn{1}{c}{ReLU} & 
            \multicolumn{1}{c}{MRI} & 
            \multicolumn{1}{C{1.1\columnwidth}}{Addressed the spatial invariance challenge caused by pooling and padding operations by using a pixel-wise 4D position vector (distance from top, right, bottom, and left) as input to train an MLP representing pixel-wise distributation in order to generate the mean and covariance matrix of that pixel.} & 
            \multicolumn{1}{lV{4}}{\cellcolor{white}\cite{zhang2021nerd}}  \\
            \cline{2-8}

            \rowcolor{llightpurple}  
            \multirow{-4}{*}{\cellcolor{lightpurple}\textbf{\scriptsize Segmentation}} & 
            \multicolumn{1}{c}{\MRT{2}{Retinal INR}} & 
            \multicolumn{1}{c}{\MRT{2}{$(x, y)$}} & 
            \multicolumn{1}{c}{\MRT{2}{RGB Value}} & 
            \multicolumn{1}{c}{\MRT{2}{\xmark}} & 
            \multicolumn{1}{c}{\MRT{2}{\Gape[0pt][2pt]{\makecell{Fundus\\Photography}}} }& 
            \multicolumn{1}{C{1.1\columnwidth}}{An INR model enhances the retinal image resolution, while a Vision Transformer (ViT) conducts self-distillation on the original image to extract the key features. These features are utilized for the segmentation of retinal vessels, therefore aiding in the detection of ocular diseases.} & 
            \multicolumn{1}{lV{4}}{\cellcolor{white}\cite{gu2022retinal}}  \\
            \hline



            \rowcolor{llightpink}  
            \cellcolor{lightpink} & 
            \multicolumn{1}{c}{\MRT{2}{MedNeRF}} & 
            \multicolumn{1}{c}{\MRT{2}{$(x, y, z, \theta, \phi)$}} & 
            \multicolumn{1}{c}{\MRT{2}{(RGB, $\sigma$)}} & 
            \multicolumn{1}{c}{\MRT{2}{ReLU}} & 
            \multicolumn{1}{c}{\MRT{2}{CT}} & 
            \multicolumn{1}{C{1.1\columnwidth}}{Combines NeRF and a CNN (inspired by GRAF~\cite{schwarz2020graf}) to generate CT projections from X-rays by training NeRF as the generator to output image patches and a CNN as the discriminator to refine NeRF outputs, enhancing image quality and addressing NeRF's struggles with complex scenes.} & 
            \multicolumn{1}{lV{4}}{\cellcolor{white}\cite{corona2022mednerf}}  \\
            \cline{2-8}

            \rowcolor{llightpink}  
            \cellcolor{lightpink} & 
            \multicolumn{1}{c}{\Gape[0pt][2pt]{\MRT{3}{\makecell{Surgical\\Neural \\Rendering}}} } & 
            \multicolumn{1}{c}{\MRT{3}{$(x, y, z, \theta, \phi)$}} & 
            \multicolumn{1}{c}{\MRT{3}{(RGB, $\sigma$)}} & 
            \multicolumn{1}{c}{\MRT{3}{ReLU}} & 
            \multicolumn{1}{c}{\Gape[0pt][2pt]{\MRT{3}{\makecell{Endoscopic \\ imaging}} }} & 
            \multicolumn{1}{C{1.1\columnwidth}}{NeRF-based rendering in robotic surgery that captures non-rigid deformations and reconstruct the 3D structures of the surgical scenes. from single-viewpoint stereo endoscopes. It handles occlusion caused by surgical instruments by utilizing a canonical radiance field and a time-dependent displacement field. The canonical field maps coordinates and viewing directions to colors and space occupancy, while the displacement field maps input space-time coordinates to displacement vectors.} & 
            \multicolumn{1}{lV{4}}{\cellcolor{white}\cite{wang2022neural}}  \\
            \cline{2-8}

            \rowcolor{llightpink}  
            \multirow{-3}{*}{\cellcolor{lightpink}\textbf{\scriptsize Neural Rendering}} & 
            \multicolumn{1}{c}{\MRT{3}{SNAF}} & 
            \multicolumn{1}{c}{\MRT{3}{$(x, y, z)$}} & 
            \multicolumn{1}{c}{\Gape[0pt][2pt]{\MRT{3}{\makecell{ Attenuation \\ Coefficient} }}} & 
            \multicolumn{1}{c}{\MRT{3}{ReLU}} & 
            \multicolumn{1}{c}{\MRT{3}{CBCT}} & 
            \multicolumn{1}{C{1.1\columnwidth}}{$\bullet$ Neural rendering is utilized as part of the CBCT reconstruction process to enhance the quality of the rendered CBCT images by learning neural attenuation fields using a multi-resolution hash table and employing volume rendering, high-quality CBCT images are generated from sparse 2D projections. \newline $\bullet$ The deblurring network takes as input a rendered novel view from the learned neural attenuation field, along with its neighboring views to mitigate the blurring effect resulting from limited input projections.} & 
            \multicolumn{1}{lV{4}}{\cellcolor{white}\cite{fang2022snaf}}  \\
            \hline


            \rowcolor{llyellow}  
            \cellcolor{lyellow} & 
            \multicolumn{1}{c}{\multirow{4}{*}{SCI}} & 
            \multicolumn{1}{c}{\Gape[0pt][2pt]{\multirow{4}{*}{\makecell{$(x, y, z)$ \\ 
            }}}} & 
            \multicolumn{1}{c}{\Gape[0pt][2pt]{\multirow{4}{*}{\makecell{Compressed \\Representation
            }}} }& 
            \multicolumn{1}{c}{\multirow{4}{*}{Sine}} & 
            \multicolumn{1}{c}{\multirow{4}{*}{CT}} & 
            \multicolumn{1}{C{1.1\columnwidth}}{$\bullet$ Mitigates INR's limitations on broad-spectrum data by introducing adaptive partitioning, which divides the data into blocks within INR's spectrum envelop and compresses each block using a funnel-shaped neural network architecture (wider beginning and narrower end) to capture its spectrum and characteristics, resulting in compressed data obtained through parameter optimization and serialization. \newline
            $\bullet$ Their findings demonstrate that INR struggles to accurately represent data with a wide range of frequencies, impacting its fidelity in capturing diverse spectral components.} & 
            \multicolumn{1}{lV{4}}{\cellcolor{white}\cite{yang2023sci}}  \\
            \cline{2-8}

            \rowcolor{llyellow}  
            \multirow{-3}{*}{\cellcolor{lyellow}\textbf{\scriptsize Compression}} & 
            \multicolumn{1}{c}{\multirow{2}{*}{\MRT{2}{TINC}}} & 
            \multicolumn{1}{c}{\MRT{3}{$(x, y, z)$ 
            }} & 
            \multicolumn{1}{c}{\Gape[0pt][2pt]{\MRT{3}{\makecell{Compressed \\Representation}}}} 
             & 
            \multicolumn{1}{c}{\MRT{3}{Sine}} & 
            \multicolumn{1}{c}{\MRT{3}{Phase-Contrast CT}} & 
            \multicolumn{1}{C{1.1\columnwidth}}{Uses octree partitioning to enable visually similar blocks to share parameters within a tree-shaped neural network structure, enhancing representation compactness and resulting in a more efficient and concise architecture, achieved through an MLP-based implicit neural function representation of each equal-sized block of target data.} & 
            \multicolumn{1}{lV{4}}{\cellcolor{white}\cite{yang2023tinc}}  \\
            \hline


            \rowcolor{lskyblue}  
            \cellcolor{skyblue} & 
            \multicolumn{1}{c}{\MRT{2}{IDIR}} & 
            \multicolumn{1}{c}{\MRT{2}{$(x, y, z)$}} & 
            \multicolumn{1}{c}{\MRT{2}{Deformation Vector}} & 
            \multicolumn{1}{c}{\MRT{2}{Sine}} &
            \multicolumn{1}{c}{\MRT{2}{CT}} & 
            \multicolumn{1}{C{1.1\columnwidth}}{Proposes an alternative approach to image registration by optimizing a MLP instead of a CNN. The authors leverage insights from differentiable rendering to demonstrate how an implicit deformable image registration  model can be combined with regularization terms using automatic differentiation techniques. } & 
            \multicolumn{1}{lV{4}}{\cellcolor{white}\cite{wolterink2022implicit}}  \\
            \cline{2-8}

            \rowcolor{lskyblue}  
            \multirow{-1}{*}{\cellcolor{skyblue}\textbf{\scriptsize Registration}} & 
            \multicolumn{1}{c}{\multirow{3}{*}{mirnf}} & 
            \multicolumn{1}{c}{\multirow{3}{*}{$(x, y, z)$}} & 
            \multicolumn{1}{c}{\Gape[0pt][2pt]{\multirow{3}{*}{\makecell{Displacement Vector \\ 
                        Velocity Vector}}}} & 
            \multicolumn{1}{c}{\MRT{3}{Sine}} & 
            \multicolumn{1}{c}{\MRT{3}{MRI}} & 
            \multicolumn{1}{C{1.1\columnwidth}}{A novel framework that combines optimization with deep neural networks for image registration. It utilizes neural fields to represents the transformation between pair of images, offering two methods for generating deformation fields. The optimal registration is achieved through parameters of neural field update via stochastic gradient descent. The paper also discusses ways to enhance model optimization.} & 
            \multicolumn{1}{lV{4}}{\cellcolor{white}\cite{sun2022mirnf}}  \\
            
             \hlineB{4}
             
        \end{tabular}
    }
\end{table*}
\endgroup

\section{Comparative Overview}\label{sec:comparative_overview}

To provide a comparative overview, we have organized comparative information and findings in the \autoref{tab:highlights}. According to the table, it is evident that image reconstruction has attracted more interest than tasks like segmentation, compression, registration, and others. This preference is mainly driven by their great ability to enhance resolution and reduce noise, especially in medical scenarios where the imaging device is prone to uncertainty. We discuss and compare noteworthy elements in the following:

\noindent\textbf{Defining Parameters:} The parameters used as the input to INR are not always cartesian coordinates and depend on the task and the signal distribution that the neural network is defining. For instance, CoiL~\cite{sun2021coil} tried to define the measurement field by using the parameters that characterize a sensor response including the viewing angle and spatial location of the detector. Likewise, NeRD~\cite{zhang2021nerd} used the positional distance in cardinal directions to define the pixel-wise distribution function. 

\noindent\textbf{Local Information:} 
It is worth noting that the methods employing CNNs, such as ArSSR~\cite{wu2022arbitrary}, BS-ISR~\cite{barrowclough2021binary}, and MedNeRF~\cite{corona2022mednerf}, use specifically leverage the power of CNNs to incorporate local semantic information during the representation process. By utilizing the convolutional layers, these methods can capture and encode local features and spatial relationships, enabling more accurate and context-aware representations for tasks such as noise removal, boundary modeling, and super-resolution.

\noindent\textbf{Sparse View CT Reconstruction:} 
As aforementioned in \autoref{sec:importance}, reducing exposure of patients to radiation dose plays a significant role in improving health care systems. As a result, a notable number of works have developed various strategies to reconstruct CT images with sparse and limited measurements and projection data. Both NeRP~\cite{shen2022nerp} and CoiL~\cite{sun2021coil} address the challenge of sparse CT reconstruction by leveraging prior information or geometric relationships. DCTR~\cite{reed2021dynamic} addresses this challenge in the context of dynamic 4D-CT reconstruction, which is suited for moving structures, such as organs affected by respiration or cardiac motion. In cone-beam computed tomography (CBCT), only the area of interest is exposed to radiation, which reduces radiation exposure to surrounding tissues and organs. SNAF~\cite{fang2022snaf} studied reconstructing scans for this special medical imaging technique by utilizing a neural rendering method to implicitly learn an attenuation field. However, due to the use of limited input projections, the resulting outputs are blurry and require additional effort. It's important to note that sparse view reconstruction is a technique that trades off radiation dose reduction with the potential loss of image quality and accuracy, which is why INR gathered a lot of research attention in this field.

\noindent\textbf{Network type: SIREN-based vs NeRF-based:} Most of the works reviewed are using ReLU MLPs with Fourier mapping applied to their input to mitigate spectral bias. Since neural volume rendering is based on NeRF~\cite{mildenhall2021nerf} design for view synthesis and continuous representation, the activation function is ReLU with Fourier features as input to model 3D structure of the scene accurately~\cite{corona2022mednerf,wang2022neural,fang2022snaf}. Nonetheless, volume rendering in medical scenarios differs in terms of the surface boundary, as the entire organ holds valuable diagnostic information compared to other domains using NeRF. The type of network is influenced by the objective of the task it aims to solve. Higher-order differentiability of periodic activations enables incorporating more advanced regularization terms into the optimization process of the registration, such as Jacobian regularizer, hyperelastic regularizer~\cite{burger2013hyperelastic}, and bending energy penalty~\cite{rueckert1999nonrigid}, which is used in IDIR~\cite{wolterink2022implicit} method.

\section{Future Work and Open Challenges}\label{sec:results}
Despite the benefits of INRs, particularly in the field of medicine, they are still limited in various aspects and require further research efforts to become viable for practical applications, given the high-stakes nature of the medical domain. We discuss these limitations briefly in the following.

\noindent\textbf{Computational complexity and training time:} 
Learning a neural representation for each signal separately involves a considerable amount of memory and computational resources. Furthermore, fitting an INR for applications involving high-dimensional data like 3D volumes can be time-consuming~\cite{saragadam2023wire}. This can pose challenges for real-time applications that require immediate responses. The complexity arises from factors such as the size of the input data and the model architecture. Meta-learning and multi-scale representations help accelerate training time and optimize memory utilization in several domains \cite{gao2023sinco,saragadam2022miner}, which provide pathways for representing anatomical and biological structures with reduced training time and greater practicality.


\noindent\textbf{Scaling to more complex signals:} 
To better represent higher-resolution signals or complex 3D shapes with fine detail can be challenging. The mapping involved in such representations is often highly nonlinear, making it difficult to scale up without incurring significant computational costs. Both widening and deepening the MLP can enhance its representation capability, but the backpropagation algorithm used for training deep neural networks becomes more computationally intensive as the depth increases, and the vanishing/exploding gradient problem may arise. Researchers often need to strike a balance between model complexity and available computational resources. Various techniques have been developed \cite{chen2023mobilenerf,mueller2022instant,kadarvish2021ensemble}.

\noindent\textbf{Video-based INR:} When it comes to decoding time, video compression methods employing INR is better compared to other models \cite{chen2021nerv}. This functionality allows parallel processing in feed-forwarding, enabling the independent computation of each frame during decoding. As a result, they got the most attention in robotic-assisted surgery where both speed and accuracy are critical \cite{schmidt2022recurrent,schmidt2022fast,wang2022neural}. However, modeling semantic relationships between frames in high-frequency videos (i.e., high frame rate) presents considerable challenges~\cite{zhao2023dnerv}, and ongoing research and development are crucial to fully utilize the potential of INRs in this field.

\section{Conclusion}\label{sec:results}
In conclusion, this survey has offered a comprehensive overview of INRs within the realm of medical imaging. Through the utilization of neural networks and implicit continuous functions, INRs have demonstrated substantial potential in tackling complex issues within medical settings. The survey has emphasized the benefits of employing INRs and has delved into their application across various medical imaging tasks. Additionally, it has identified open challenges and areas of future research, providing valuable insights for researchers in the field.
{\small
\bibliographystyle{ieee_fullname}
\bibliography{egbib}
}

\end{document}